\documentclass[conference]{IEEEtran}
\IEEEoverridecommandlockouts    

\usepackage{cite}
\usepackage{acro}
\DeclareAcronym{BET}{short = BET, long = battery electric truck, short-plural = s, long-plural = s}
\DeclareAcronym{FRLM}{short = FRLM, long = flow-refueling location model, short-plural = s, long-plural = s}
\usepackage{amsmath,amssymb,amsfonts}
\usepackage{algorithmic}
\usepackage{graphicx}
\usepackage{textcomp}
\usepackage{xcolor}
\usepackage{cuted}
\usepackage{flushend}
\usepackage{makecell}
\usepackage{acro}
\usepackage{multirow}

\title{\LARGE \bf Multi-Phase Optimization of Shared Charging Infrastructure for Freight Electrification}

 \author{
 	\parbox{\textwidth}{%
 		\centering
 		Joas Kahlert$^{1}$, Jiali Fu$^{2}$, Chengxi Liu$^{3}$%
 	}%
 	\thanks{$^{1}$Decision and Control Systems, Royal Institute of Technology (KTH), Stockholm, Sweden
 		{\tt\small kahlert@kth.se}}%
 	\thanks{$^{2}$Swedish National Road and Transport Research Institute (VTI), Stockholm, Sweden
 		{\tt\small jiali.fu@vti.se}}%
    \thanks{$^{3}$Swedish National Road and Transport Research Institute (VTI), Stockholm, Sweden
 		{\tt\small chengxi.liu@vti.se}}%
 }

\begin{document}
	
\maketitle
\thispagestyle{empty}
\pagestyle{empty}

\begin{abstract}
The transition to heavy-duty battery electric vehicles requires an efficient and cost-effective deployment of the charging infrastructure, particularly when multiple operators share resources. This paper presents a multi-phase optimization framework for the joint planning of charging stations in a shared network, using high-resolution empirical truck trajectory data from two freight companies with distinct operational characteristics. The model is formulated to minimize the total number of charging stations while ensuring that the predefined electrification targets are met over successive expansion stages.

The analysis captures heterogeneity in fleet usage, with one company operating a spatially concentrated network with shorter and more consistent routes, and the other exhibiting more dispersed operations with longer and more variable driving patterns. The results show that early-stage infrastructure deployment primarily supports fleets with concentrated operations, while later expansion phases are essential to accommodate long-haul and geographically dispersed transport demand. Furthermore, shared infrastructure not only enables reductions in redundant investments, but also introduces dependencies where certain fleets rely heavily on the full network to sustain electrified operations.

In general, the findings highlight the importance of coordinated and data-driven infrastructure planning, and demonstrate that fleet-specific characteristics strongly influence both infrastructure requirements and electrification outcomes. The proposed framework provides practical insights on how collaborative and phased deployment strategies can enhance the scalability and efficiency of freight transport electrification.

\end{abstract}

\begin{center}
\small
This work has been submitted to the IEEE for possible publication. 
Copyright may be transferred without notice, after which this version may no longer be accessible.
\end{center}
\section{Introduction}
\label{sec:introduction}
The freight transport sector is undergoing rapid transformation driven by increasing environmental concerns, regulatory pressures, and the need to reduce greenhouse gas emissions. In this context, the electrification of heavy-duty transport through \acp{BET} has emerged as a key strategy to achieve sustainable logistics operations. However, despite ongoing technological advancements, the large-scale adoption of \acp{BET} remains constrained by economic, operational, and infrastructure challenges.

One of the critical barriers is the deployment of an adequate charging infrastructure. Unlike conventional refueling systems, electric charging networks require careful planning due to high installation costs, limited grid capacity, and the need to align infrastructure placement with vehicle routing patterns. These challenges are further amplified in freight transport, where operational constraints such as long driving distances, heterogeneous fleet use, and tight delivery schedules must be considered. In particular, variability in daily driving distances and spatial distribution of routes significantly affects the feasibility of electrification, as vehicles must reliably access charging facilities without disrupting logistics operations.

To address these challenges, recent research has emphasized data-driven infrastructure planning based on real-world transport patterns. However, most studies adopt a single-operator perspective or assume fully developed networks. In practice, freight electrification is gradual, with infrastructure deployed incrementally and often shared among stakeholders to reduce costs and improve utilization.

This study contributes by proposing a multi-phase optimization framework for shared charging infrastructure deployment among freight operators. Using high-resolution data from two Swedish logistics companies, the approach captures heterogeneity in fleet operations, including route length, spatial distribution, and daily variability. The model determines the minimum number of charging stations required to meet electrification targets across successive deployment phases, reflecting a realistic stepwise transition. The key contribution lies in analyzing the interaction between shared infrastructure and phased deployment under heterogeneous fleet conditions. By modeling incremental electrification goals and interdependencies between operators, the study provides insight into how collaborative planning can reduce redundant investments while ensuring adequate coverage. The results support decision-makers in designing cost-effective and scalable charging networks for real-world freight operations.

\section{Related Research}
The transition to electric freight transport has stimulated extensive research on the planning and deployment of charging infrastructure for \acp{BET}. A foundational framework in this domain is \ac{FRLM}, introduced by \cite{Kuby2005, Kuby2007}, which determines optimal station locations such that vehicle trips remain feasible under range constraints. This formulation is particularly relevant for freight applications, where vehicles operate in tour-based patterns and must complete sequences of trips within battery limitations. However, classical implementations of \ac{FRLM} typically rely on predefined routes and deterministic assumptions, which may limit their applicability in real-world logistics environments.

To address these limitations, recent studies have incorporated additional operational and system-level considerations. In particular, integrated approaches jointly optimize vehicle routing and charging infrastructure while accounting for constraints in power systems and vehicle operations. For example, data-driven models based on truck trajectory data have been proposed to identify strategic charging locations that reflect actual freight movement patterns \cite{Li2025}. Similarly, studies using multi-day transport data highlight the importance of temporal variability and spatial concentration of freight demand when designing charging networks \cite{Fu2024}. Furthermore, a recent study  \cite{Ingelstrom2026} emphasizes the role of intelligent and data-driven charging infrastructure design, highlighting how optimization and forecasting techniques can improve infrastructure utilization and system performance under uncertain demand conditions. These approaches underline the importance of combining operational data with infrastructure planning models.

Another important research direction focuses on infrastructure planning under uncertainty and over multiple time periods. Multi-period and stochastic optimization models have been developed to capture the gradual adoption of \acp{BET} and the evolving demand for charging services \cite{Pagnier2026}. A recent work on ramp-up strategies explicitly considers the sequential deployment of charging infrastructure, demonstrating how early-stage investment decisions influence long-term network performance and electrification feasibility \cite{GEORG2024}. Such approaches emphasize the need for flexible and adaptive planning strategies, particularly in early-stage electrification scenarios.

Comprehensive review studies further underline the multi-dimensional nature of charging infrastructure planning. These surveys emphasize that effective solutions must simultaneously consider spatial demand distribution, grid integration, user behavior, and economic factors \cite{Ullah2024, Ma2022, Unterluggauer2022}. Despite these advances, most existing work focuses on single-operator settings or long-term optimal configurations, with limited attention to shared infrastructure scenarios, ramp-up strategies, and early-stage deployment dynamics.

\section{Methodology}
\label{sec:designandimplementation}

\subsection{Problem Specification}
In our previous work \cite{Kahlert2025}, we studied a charging station placement problem in a multi-phase expansion framework. In each phase, a limited number of charging stations are added to an existing network in a stepwise manner, referred to as a step-up process. The objective in each phase is to maximize the fraction of electrifiable traffic given the current charging infrastructure and the strategic placement of new stations. For a detailed description of the formulation and methodology of the problem, we refer to \cite{Kahlert2025}.

Building on this multi-phase expansion approach, we extend the framework to incorporate shared charging infrastructure across multiple fleet operators. From this multi-fleet perspective, the objective is to minimize the number of charging stations required to achieve predefined electrification thresholds for all fleets. By assigning fleet-specific thresholds, the model enables control over fairness among operators. In this way, fleet electrification progresses at prescribed rates over successive expansion phases, rather than favoring fleets with more advantageous operational characteristics.

\subsection{Data Description and Preprocessing}

The dataset used in this study comprises real-world freight operation data collected from approximately 200 vehicles operated by two logistics companies operating in Sweden, hereafter referred to as Company A and Company B. The data is organized into two primary types of telemetric events:
\begin{itemize}
\item Start and stop events are recorded whenever the vehicle control unit detects ignition activation or shutdown. 
\item Precision events are recorded at regular intervals of approximately 10–30 seconds during vehicle operation.
\end{itemize}

Each recorded event includes a timestamp, geographic coordinates, and a unique vehicle identifier. When chronologically ordered, these data provide a detailed representation of individual vehicle trajectories and allow the reconstruction of spatiotemporal movement patterns across the transport network. The data collection spans from November 2021 to March 2024 without significant interruptions, resulting in a large-scale dataset. In total, the dataset contains about 60 million precision events and 3 million start/stop events, corresponding to a cumulative driven distance of 2.2 million kilometers.

The preprocessing step of the data used in this problem is extensive and is described in detail in \cite{Kahlert2025} along with insightful visualizations. In short, the preprocessing procedure maps the telemetric truck trajectories to a network of nodes that represent frequently visited stopping locations. Applying this mapping individually to each truck yields spatiotemporal paths that describe the movement of all trucks within the transport network over a given time horizon. Such datasets are valuable for a wide range of transport analysis applications.

For the purpose of this study, we are interested in the overlap of operational characteristics between different fleets, including shared geographical regions and similarities in operational density within the transport network. Detecting common nodes across fleets can reveal promising candidate locations for shared charging infrastructure, thereby enabling more efficient resource utilization across multiple operators.

\subsection{Mathematical Model}

The problem is formulated as a Mixed Integer Linear Program (MILP). Consider a charging network with nodes $i\in\mathcal I$. A set of trucks $n\in\mathcal N_h$ belonging to the fleet $h\in\mathcal H$, operate within this network. Each truck is characterized by a battery capacity $E_n$, and follows a predetermined route through the network. A route is represented by the index set $j\in\mathcal J_n$, where each element $j=(i, i')$ corresponds to a directed segment from the origin node $i$ to the destination node $i'$, i.e., an origin–destination (OD) pair. For each pair of OD, the travel distance and associated energy consumption are precomputed and denoted by $d_j$ and $e_j$, respectively.

Consequently, the decision variables are defined as two binary indicators and one continuous variable: $y_i$, indicating whether a charging station is installed at node $i$; $z_n$, indicating whether the truck $n$ is operated as an electric vehicle; and a continuous variable $e_j$, representing the state of charge of the trucks along its route.

Using standard MILP notation and omitting nonessential details for brevity, the problem is formulated as follows:
\begin{subequations}
    \label{eq:model}
    \begin{align}
        \min_{y,z,e} \quad\ & \sum_{i\in\mathcal I}y_i-f_\varepsilon,\label{model:obj}\\ 
        \text{s.t.}\quad &\sum_{n\in\mathcal N_h}\sum_{j\in\mathcal J_n}d_jz_n\geq \beta_s \sum_{n\in\mathcal N_h}\sum_{j\in\mathcal J_n}d_j,\label{model:fairness}\\
        &e_{n,j+1}  \leq e_{n,j} - z_nd_j + y_iE_n,\label{model:energy}\\
        &0\leq e_{n,j}\leq E_n,\label{model:bounds}\\
        &y_i=1,\ z_n=1,\quad(\forall i,n\in\{\text{pre-fixed infrastructure}\})\label{model:preset}\\
        &y_i, \ z_n\in \{0,1\},\ e_{n,j}\in\mathbb{R}.\label{model:domains}\\
        &\forall i\in\mathcal I,\ \forall j\in\mathcal J_n,\ \forall n\in\mathcal N_h,\ \forall h\in\mathcal H.
    \end{align}
\end{subequations}

Equation \eqref{model:obj} minimizes the number of charging stations required. However, multiple configurations of charging stations may satisfy the constraints while achieving the same minimum. To distinguish between such equivalent solutions, the objective is enhanced with an additional term, denoted by $f_\epsilon$. This term is chosen such that it does not affect the primary optimality criterion (i.e., $f_\epsilon < 1$), but instead acts as a secondary selection mechanism among equivalent solutions. Specifically, $f_\epsilon$ is defined as a function of the desired notion of solution quality. In this study, we define it in terms of the achieved fraction of electrification.

Equation \eqref{model:fairness} ensures that each fleet operator included in the problem achieves at least a fraction $\beta_s$ of electrification across its entire fleet at step $s$ of the expansion process. Equation \eqref{model:energy} captures the charging and discharging dynamics of each truck along consecutive route segments $j,j+1$. Discharging occurs only when a truck is electric; thus, the constraint is trivially satisfied for non-electric trucks. Equation \eqref{model:bounds} restricts the state of charge by the battery capacity of each individual truck. Equation \eqref{model:preset} fixes all pre-existing infrastructure established prior to step $s$, corresponding to the outcomes of the previously computed optimal expansion steps in the process.

The multiphase expansion process is determined by iteratively solving the model \eqref{eq:model} for each step $s$. After each iteration, the electrification threshold $\beta_s$ is updated and the corresponding optimal infrastructure decisions are carried forward to subsequent steps.

\section{Results and Analysis}
\label{sec:results}

\begin{figure}[ht]
    \centering
    \includegraphics[width=1\linewidth]{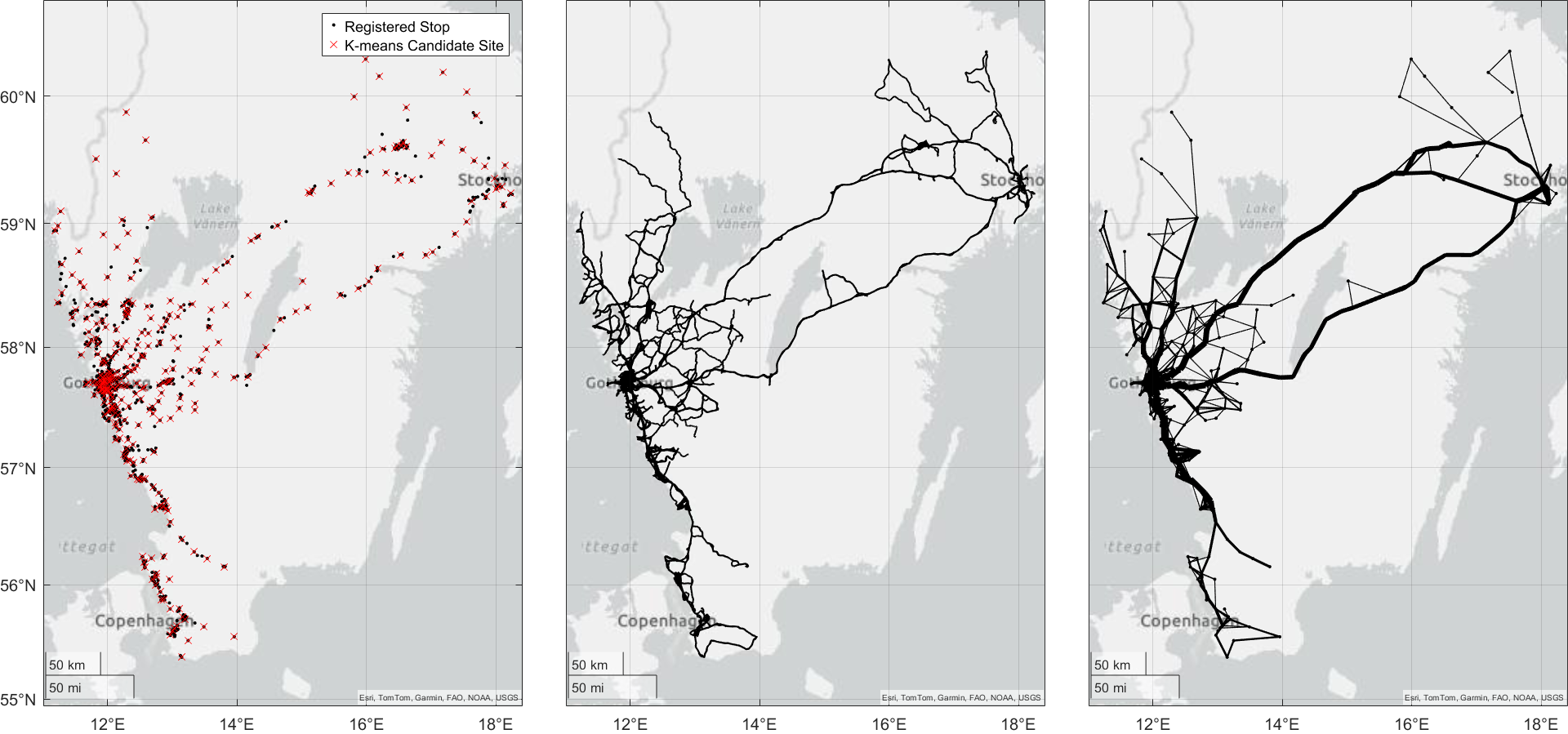}
    \includegraphics[width=1\linewidth]{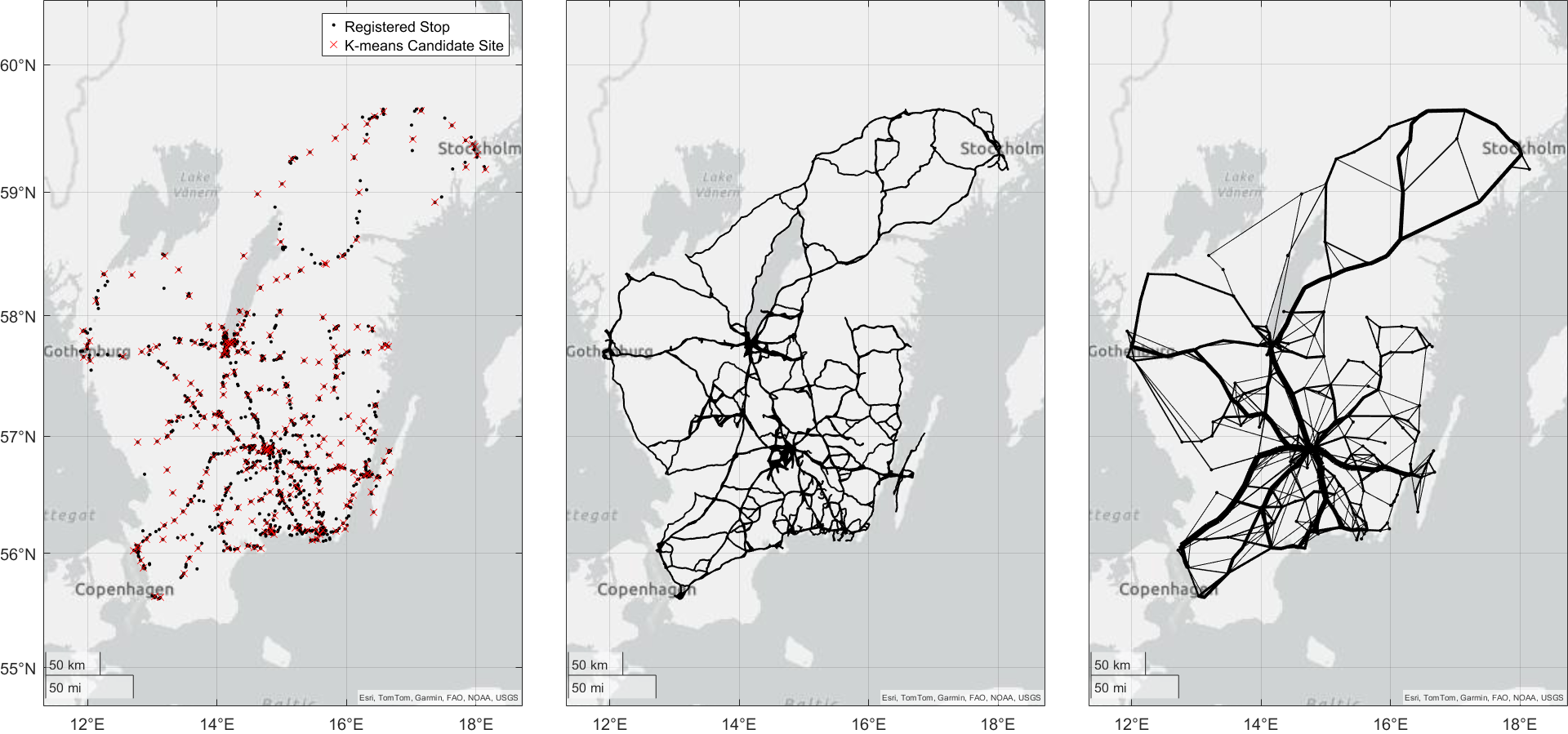}
    \caption{Data transformation and visualization for the period 2024-01-01 to 2024-03-01. Results are shown for Company A (top row) and Company B (bottom row). The left column illustrates registered stop locations (as black dots) together with the $300$ k-means cluster centroids (as red dots) representing candidate sites. The middle column displays the recorded precision events. The right column presents the aggregated network routes obtained through the data transformation process, where thicker edges indicate higher traffic intensity along the corresponding segments.}
    \label{fig:inputdata}
\end{figure}
\begin{figure}[htbp]
\centering
\hspace*{-0.5cm}
\includegraphics[width=1.1\linewidth]{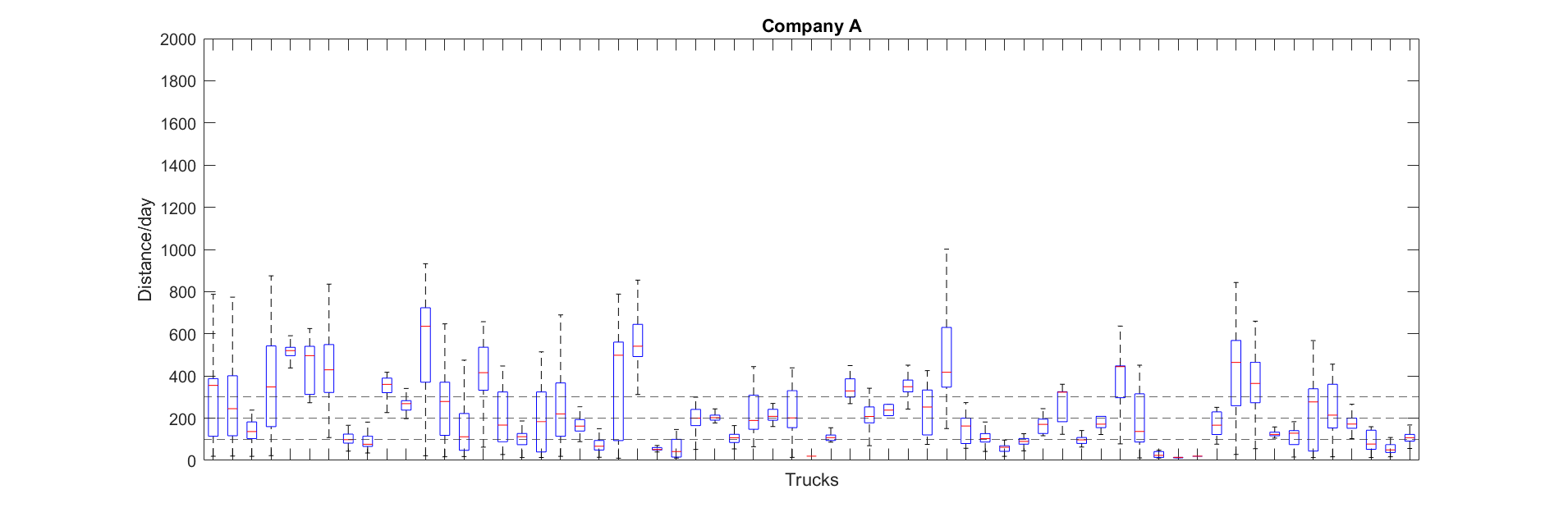}
\hspace*{-0.5cm}
\includegraphics[width=1.1\linewidth]{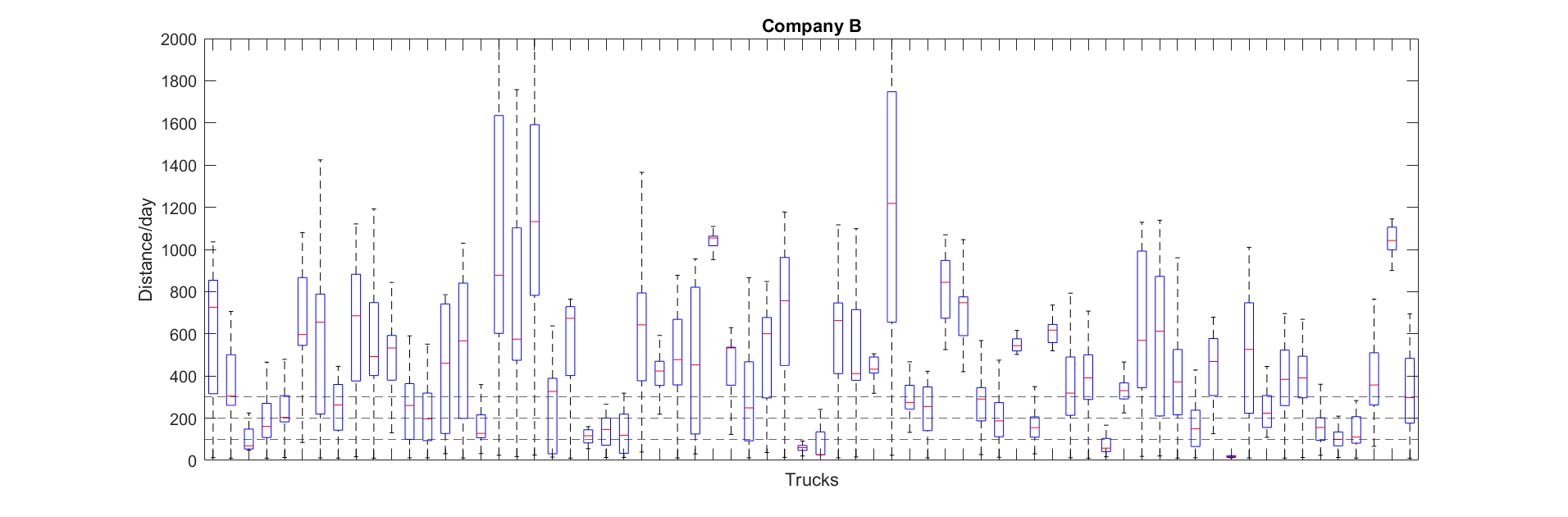}
\caption{Daily driving distances of trucks in the dataset. Each boxplot represents a truck; the red line indicates the median, and the box bounds show the 25th and 75th percentiles. Horizontal lines denote battery range classes $c_k=100,200,300$ km.}
\label{fig:boxplot}
\end{figure}

The analysis covers the period from 2024-01-01 to 2024-03-01. After preprocessing, the transport network comprises 300 nodes (see Fig.~\ref{fig:inputdata}). The dataset includes 67 trucks from Company~A and 71 from Company~B. The average daily distances are 220~km and 280~km, respectively. Figure~\ref{fig:boxplot} shows the daily distance distribution for all vehicles.

For Company~A (upper panel of Fig.~\ref{fig:boxplot}), most vehicles operate over short (around 100~km) or medium (100–300~km) distances, with few exceeding 300~km. The variability is relatively low, indicating consistent use. In contrast, Company~B (lower panel) shows a different pattern: most vehicles travel long distances ($>$300~km), with only a few in shorter ranges, and significantly higher variability in daily usage.

The map-matching procedure produces network routes that range from 100 to 1,000 nodes in length, depending on the total distance traveled by each truck. Route discontinuities occur in fewer than $<1\%$ of the cases on average. All experiments are conducted on a standard laptop equipped with an Intel i7-8550U processor running at 1.80~GHz, using Gurobi Optimizer v12.0.0 \cite{gurobi}. Solver runtimes range from a few minutes to several hours and increase with larger step sizes $Y^{max}_{s+1}-Y^{max}_s$.

The optimization problem aims to minimize the number of charging stations required in a shared network, subject to both companies achieving a specified electrification threshold. The step-up process is defined by discrete electrification progress intervals, reflecting the shared infrastructure strategy. Consequently, the electrification threshold at step $P^e_s=0.03,0.07,0.15,0.30$ for steps 1 through 4, respectively.

\begin{figure*}[htbp]
\centering
\includegraphics[width=0.9\linewidth]{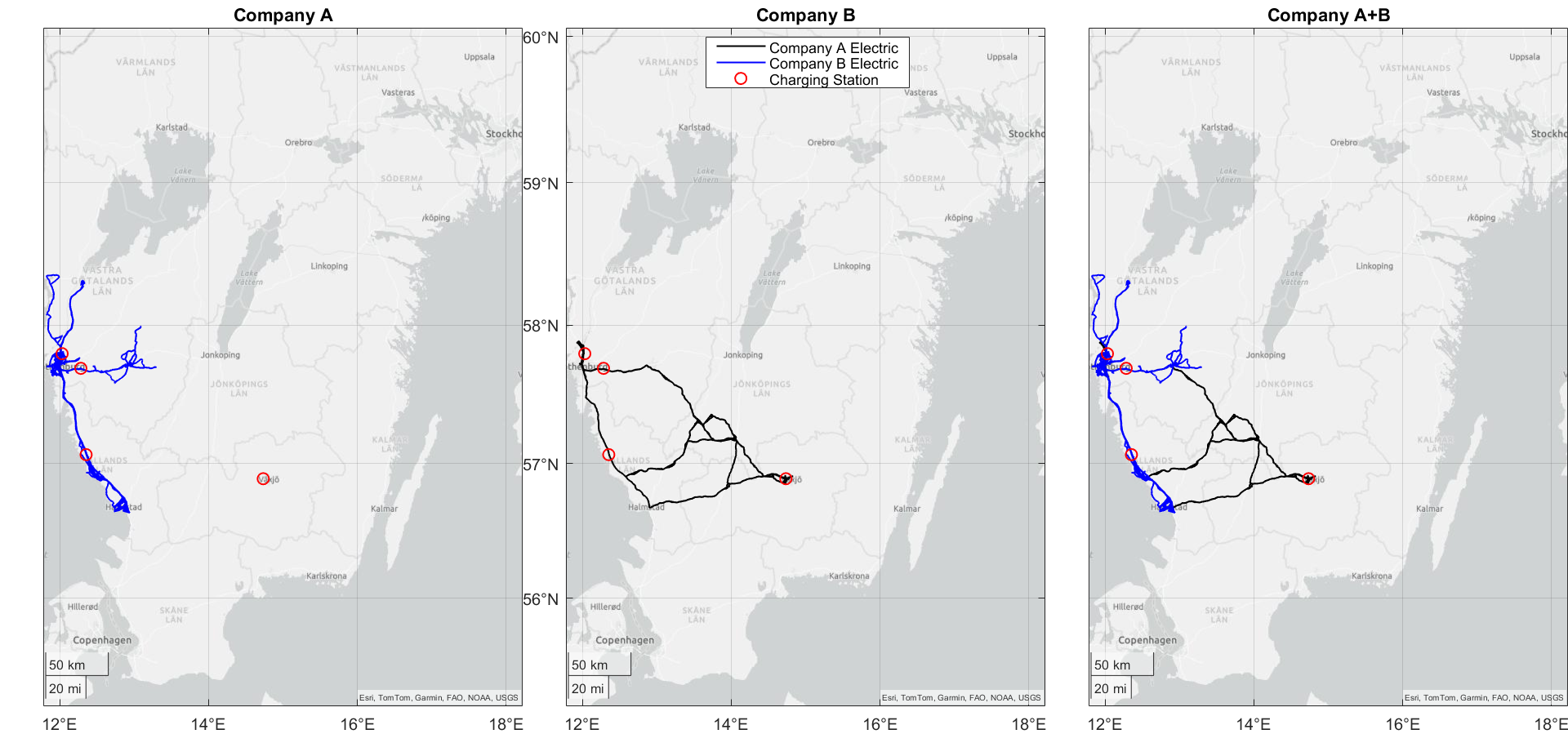}
\caption{Visualization of the optimal solution for the first expansion step with a predefined electrification threshold of $P^e_s=0.03$. The left panel depicts the optimal charging-station locations (red circles) and the electrified road network for Company A. The middle panel shows the optimal charging-station locations for Company B, while the right panel provides a combined representation of both configurations.}
\label{fig:comb_1}
\end{figure*}

Figure~\ref{fig:comb_1} shows the optimal charging station locations (red circles) in the first expansion step for an electrification threshold of $P^e_s = 0.03$. Four stations are required to achieve at least $3\%$ electrification for both companies. Three are located along the west coast near the main depot of Company~A, reflecting its concentrated network and high traffic density. This configuration yields $11.2\%$ electrification for Company~A (left panel), with $6$ of $58$ vehicles electrified using $3$ stations. The remaining station is placed at the main depot of Company~B. Due to its dispersed network, Company~B achieves only $3.2\%$ electrification (middle panel), corresponding to $1$ of $66$ vehicles, which uss all four stations. The right panel shows the combined network, and Table~\ref{tab:combsol} summarizes the results.

\begin{figure*}[htbp]
\centering
\includegraphics[width=0.9\linewidth]{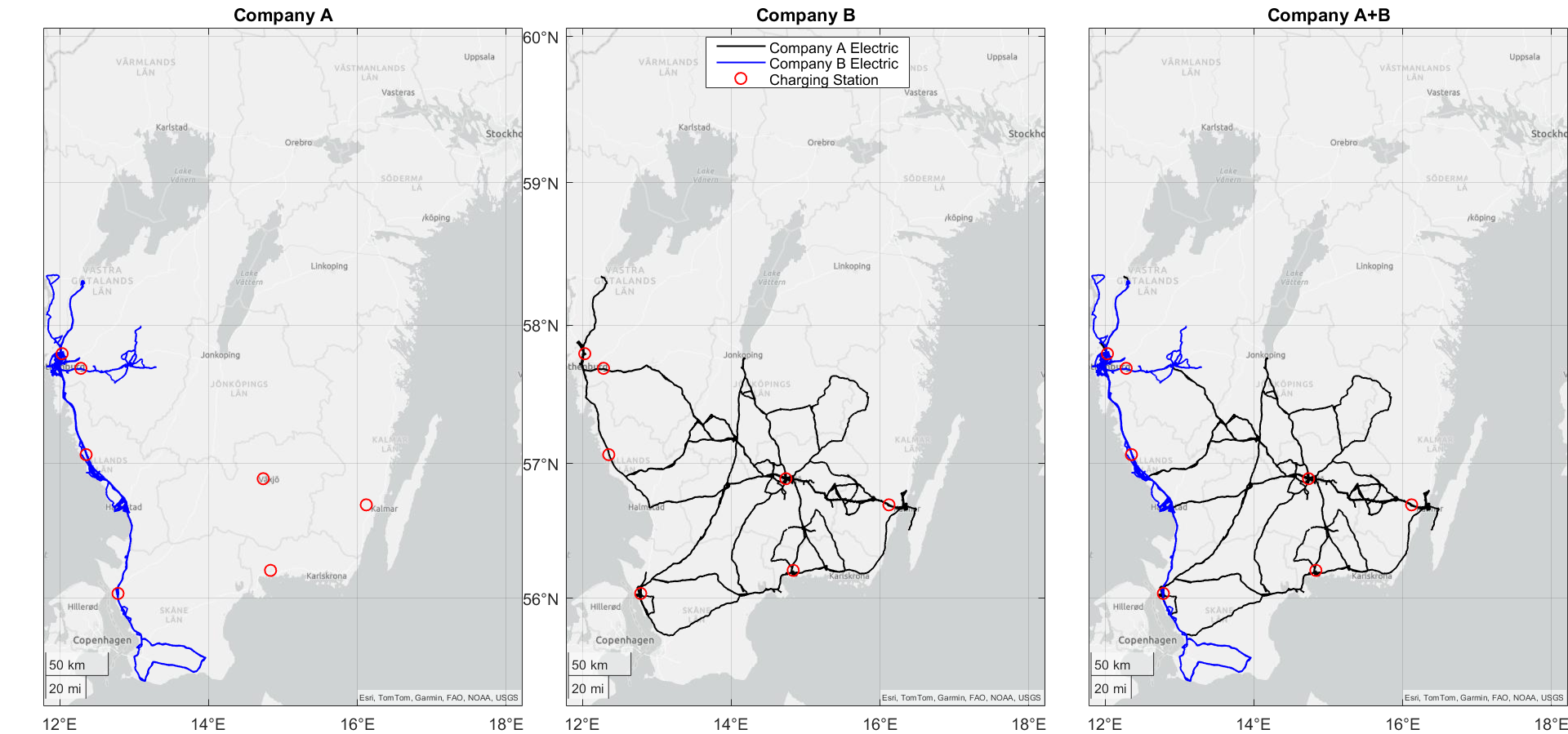}
\caption{Visualization of the optimal solution for the second expansion step with a predefined electrification threshold of $P^e_s=0.07$.}
\label{fig:comb_2}
\end{figure*}

In the second expansion step, the electrification threshold increases to $P^e_s = 0.07$, and the corresponding results are presented in Fig.~\ref{fig:comb_2}. In this phase, three additional charging stations are deployed further south. One of these stations is located along the west coast, allowing Company~A to extend its electrified transport network further south along its main transport corridor. As a result, Company~A achieves an electrified network coverage of $14.6\%$, shown in the left panel of Fig.~\ref{fig:comb_2}. In this configuration, $7$ of the $58$ vehicles are electrified and Company~A uses $4$ of the $7$ available charging stations. The remaining two charging stations are allocated further south and southeast, enabling Company~B to achieve $8.6\%$ electrification ($7$ of $66$ vehicles). As before, Company~B’s electrified routes require all available stations.

\begin{figure*}[htbp]
\centering
\includegraphics[width=0.9\linewidth]{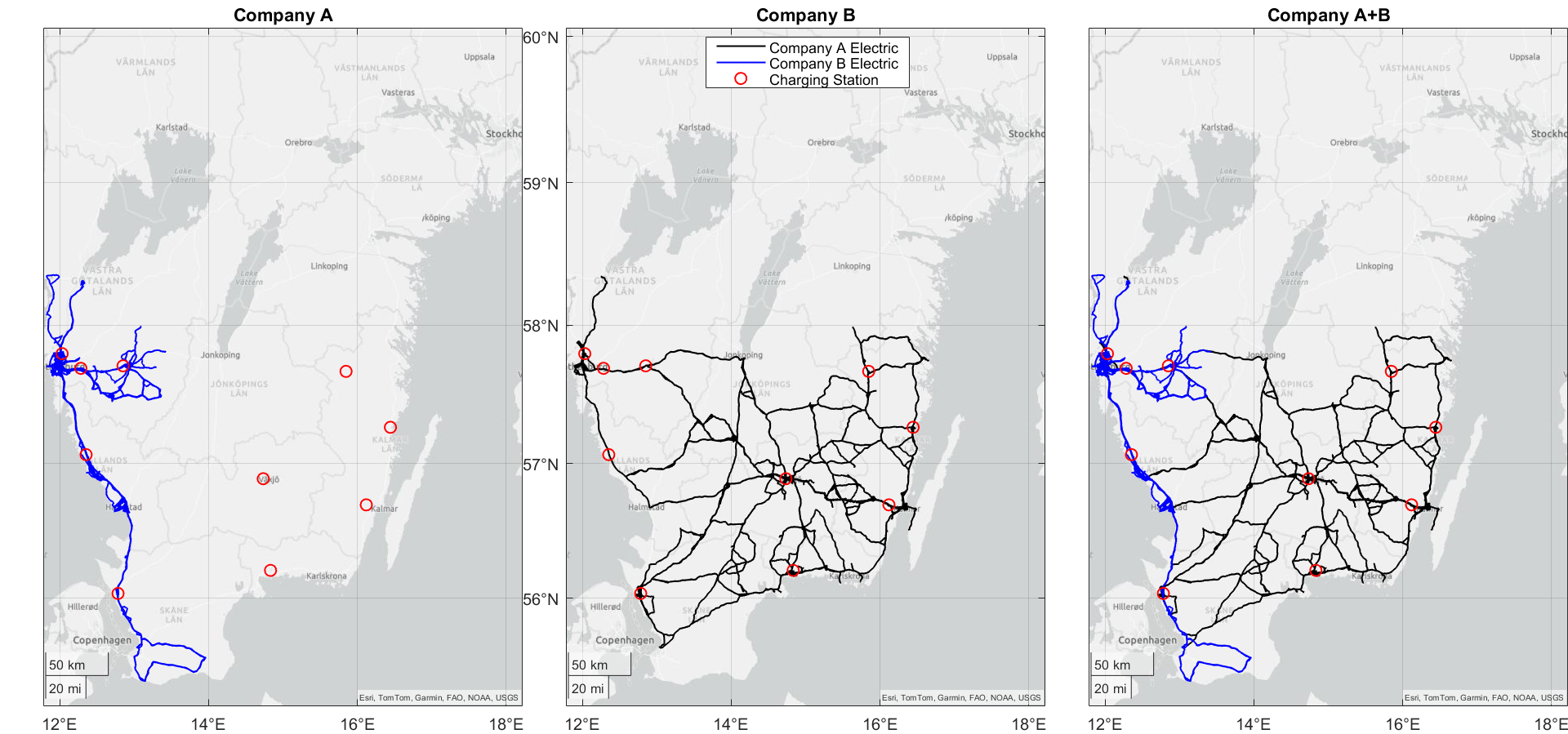}
\caption{Visualization of the optimal solution for the third expansion step with a predefined electrification threshold of $P^e_s=0.15$.}
\label{fig:comb_3}
\end{figure*}

Fig. \ref{fig:comb_3} presents the results of the third expansion step, where the electrification threshold is increased to $P^e_s = 0.15$. Three additional charging stations are deployed. One is placed east of Company~A’s main home depot (left panel in Fig.~\ref{fig:comb_3}), enabling further expansion of electrified deliveries in that area. In this configuration, Company~A uses $5$ of the $10$ available charging stations. The other two stations are located northeast of the previously installed ones (right panel in Fig.~\ref{fig:comb_3}), allowing Company~B to extend its electrified deliveries toward Stockholm and utilize all available stations. At this stage, Company~A and Company~B achieve electrified network coverages of $16.3\%$ and $18.9\%$, respectively, and both operate $8$ electric vehicles.

\begin{figure*}[htbp]
\centering
\includegraphics[width=0.9\linewidth]{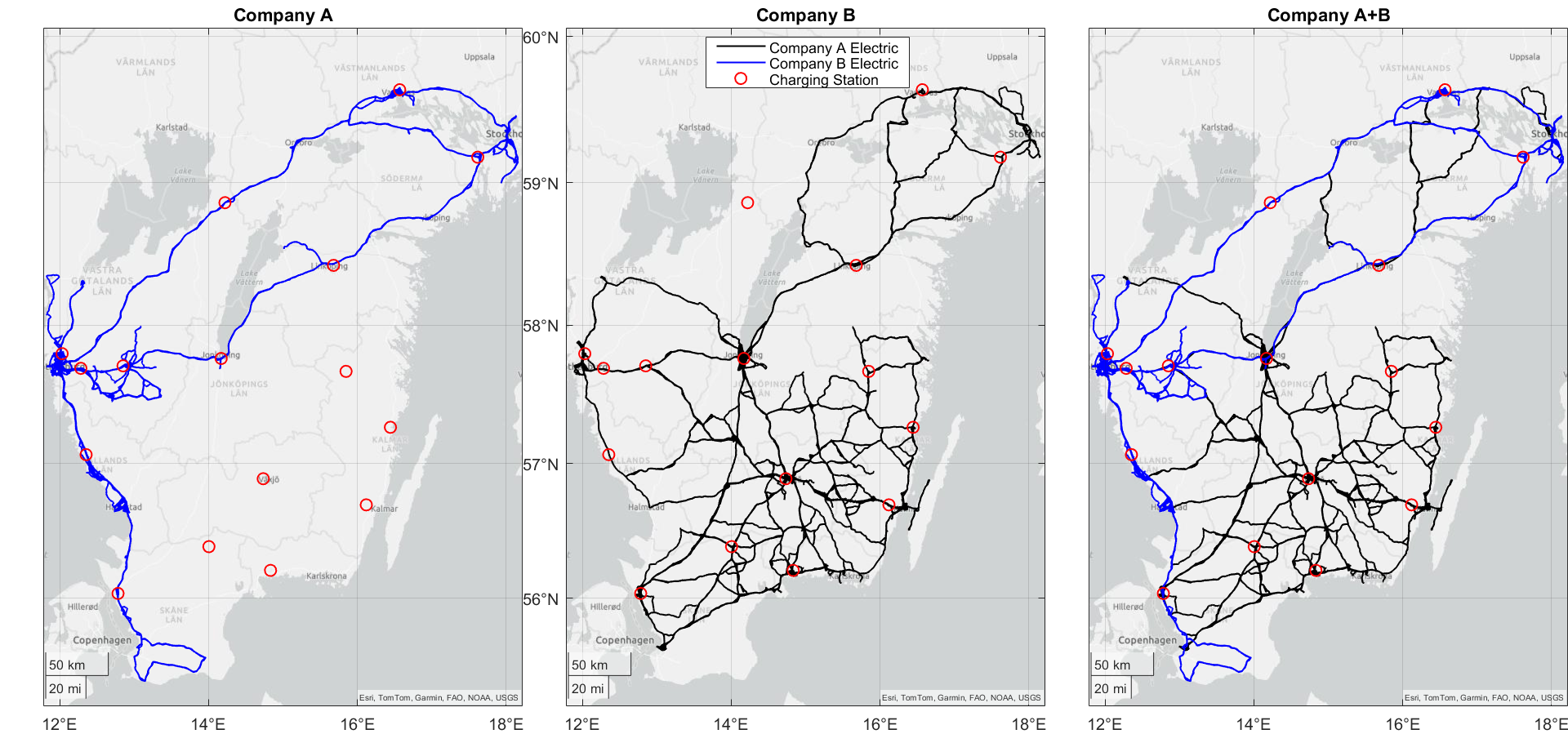}
\caption{Visualization of the optimal solution for the fourth expansion step with a predefined electrification threshold of $P^e_s=0.30$.}
\label{fig:comb_4}
\end{figure*}

In the final expansion step, the electrification threshold is increased to $P^e_s = 0.30$. Six additional charging stations are installed north of the previously deployed locations, as shown in Fig.~\ref{fig:comb_4}, enabling both companies to expand operations to Stockholm. In this configuration, Company~A achieves an electrified network coverage of $31.7\%$ with an electric fleet of $11$ vehicles, using $10$ of the $16$ available charging stations. Company~B achieves a higher coverage of $50.5\%$ with $22$ electric vehicles out of $66$ and uses all but one charging station.


\begin{table*}[htbp]
\centering
\caption{Numerical values to solutions presented in Figure \ref{fig:comb_1} to \ref{fig:comb_4}}
\begin{tabular}{c|c|c|c|c|c}
     &Step $s$&\makecell{$P^e_s$ $(\%)$}& \makecell{Fraction of \\ electric traffic $(\%)$} & \makecell{Number of \\ electric trucks} & \makecell{Number of used \\ charging stations}\\
     \hline
     \multirow{4}{*}{Company A} 
     &$1$& $3$ & $11.2$& $6/58$ &$3/4$\\
     &$2$& $7$ & $14.6$ & $7/58$& $4/7$\\
     &$3$& $15$ & $16.3$ & $8/58$& $5/10$\\
     &$4$& $30$ & $31.7$ & $11/58$ &$10/16$\\
     \hline
     \multirow{4}{*}{Company B} 
     &$1$& $3$ & $3.2$& $1/66$& $4/4$\\
     &$2$& $7$ & $8.6$ & $5/66$& $7/7$\\
     &$3$& $15$ & $18.9$ & $8/66$& $10/10$\\
     &$4$& $30$ & $50.5$ & $22/66$& $15/16$\\
     \hline 
\end{tabular}
\label{tab:combsol}
\end{table*}

\section{Summary and Conclusions}
\label{sec:conclusion}
This study investigated the optimal deployment of shared charging infrastructure to support the gradual electrification of two freight fleets operating in Sweden. Using high-resolution operational data, the analysis reveal substantially different fleet usage patterns between the two companies. Company~A operates a relatively concentrated network with lower average daily driving distances and limited variability, whereas Company~B exhibits a more geographically dispersed network with significantly longer and more variable daily routes. These structural differences play a decisive role in shaping electrification outcomes and infrastructure requirements.

The proposed optimization framework minimizes the total number of charging stations while ensuring that both companies meet predefined electrification thresholds in a stepwise expansion process. The early expansion stage primarily benefits Company~A, while later stages increasingly accommodate Company~B's long haul operations. At higher levels of electrification, Company~B achieves greater coverage of the electrified network, but depends on nearly all available charging stations, highlighting the intensity of the infrastructure of electrifying dispersed logistics networks.  

In general, the results highlight the effectiveness of the shared charging infrastructure in reducing redundant investments while accommodating diverse operational profiles. The analysis underscores that fleet characteristics—such as route length distribution, spatial concentration, and utilization variability—strongly influence both the pace and efficiency of electrification. These findings suggest that coordinated, data-driven infrastructure planning can improve the feasibility of freight electrification, particularly when multiple operators with complementary network structures collaborate. Future work may extend this framework to incorporate charging capacity constraints, temporal congestion effects, and stochastic demand to further improve real-world applicability.

\section*{ACKNOWLEDGMENTS}
The work is financially supported by the Swedish Energy Agency (Grant no. P2024-01073 and P2025-04321), and the Ministry of Science, Innovation and Universities of Spain (Grant no. PID2023-151065OB-I00). The authors wish to extend their appreciation to TRB Sweden AB, the industry collaborator who supplied us with data and offered their support and valuable feedback.

\bibliographystyle{IEEEtran}

\bibliography{root} 
	
\end{document}